\documentclass[11pt]{article}
\usepackage[margin=1in]{geometry}  
\usepackage{amsmath,graphicx,hyperref}
\usepackage{tabularx}
\usepackage{booktabs}
\usepackage{array}
\usepackage{longtable}
\usepackage{graphicx}

\title{Gaze-Aware AI: Mathematical modeling of epistemic experience of the Marginalized for Human-Computer Interaction \& AI Systems}
\author{Omkar Suresh Hatti \\ Independent Researcher \\ \texttt{omkar.hatti@icloud.com}}
\date{\today}

\begin{document}
\maketitle

\begin{abstract}
The proliferation of artificial intelligence provides an opportunity to create psychological spaciousness in society. Spaciousness is defined as the ability to hold diverse interpersonal interactions and forms the basis for vulnerability that leads to authenticity that leads to prosocial behaviors and thus to societal harmony \cite{brown2017braving,brown2012daring,foucault1977discipline, foucault1978history, foucault2003society}. This paper demonstrates an attempt to quantify \emph{gaze}, the human conditioning to subconsciously modify authentic self-expression to fit the norms of the dominant culture. Gaze is explored across various marginalized and intersectional groups, using concepts from postmodern philosophy and psychology. The effects of gaze are studied through analyzing a few redacted Reddit posts, only to be discussed in discourse and not endorsement. A mathematical formulation for the Gaze Pressure Index (GPI)-Diff Composite Metric is presented to model the analysis of two sets of conversational spaces in relation to one another. The outcome includes an equation to train Large Language Models (LLMs) - the working mechanism of AI products such as Chat-GPT; and an argument for affirming and inclusive HCI, based on the equation, is presented. The argument is supported by a few principles of Neuro-plasticity, The brain's lifelong capacity to rewire \cite{medina2014brain}. 
\end{abstract}

\textbf{Keywords:} Gaze-Aware AI, Epistemic Injustice, Human-Computer Interaction, AI Systems, Trauma-Informed AI, Social Justice, Neuro-plasticity, Inclusion, DEI

\section{Introduction}
Trigger warning: Some of these mechanisms might be probing to readers with traumatic backgrounds and it might be worth viewing them purely as \textquotedblleft mechanisms\textquotedblright{} rather than judging the ethics of it or dwelling on them. The purpose isn’t to categorize, judge, or support the morality of mechanisms but to present and understand them.

Jean-Paul Sartre’s notion of \emph{bad faith} frames the gaze as a distortion of one’s authentic self, especially when marginalized individuals interact with dominant culture norms. Sartre emphasizes that humans are not predefined like objects, they create themselves through existence. Gaze, the human agency to be a subject of one's own reality, in this context, becomes a tool of self-surveillance and conformity, the paper presents a argument further supporting this claim \cite{sartre1943being}.

Michel Foucault’s work on soft power structures like schools, hospitals, and now AI
demonstrates how norms condition behavior through institutional discipline rather than force in the modern world\cite{foucault1977discipline}.

Miranda Fricker's work introduces \emph{epistemic injustice}: testimonial injustice (when someone is discredited based on identity) and hermeneutical injustice (when someone lacks language to describe their experiences) \cite{fricker2007epistemic}. Fricker's work is further combined with MacIntyre’s emphasis on perceived virtue \cite{macintyre1981after}, we see how marginalized expressions are often diminished or misunderstood through this framework.

Judith Butler explores how gender is performed and not innate, challenging rigid binaries \cite{butler1990gender, butler1993bodies}. Monique Wittig argues that gender roles are imposed and sustained through linguistic and cultural norms \cite{wittig1992straight, wittig2007guerilleres}. Erving Goffman’s concepts of performance and stigma show how people adjust expression to manage perception \cite{goffman1959presentation, goffman1963stigma}. Gayatri Spivak discusses how marginalized voices are often unheard even when they speak \cite{spivak1988can}. Sara Ahmed offers insights into embodied emotion, social anxiety, and queer reorientation, advocating for agency in turning away from imposed norms \cite{ahmed2004cultural, ahmed2006queer}. These authors break down the term \textquotedblleft marginalized expressions\textquotedblright{} to specific indicators for modeling realities of Women, queer and persons of color (POC) identities.

Kimberl\'{e} Crenshaw’s \emph{intersectionality} explains how compounded identities multiply gaze pressure \cite{crenshaw1991mapping}. Finally, bell hooks emphasizes that love and authenticity are radical acts in oppressive environments to emphasizing urgency of affirming HCI design. \cite{hooks1981aint, hooks2000feminist}. 

Together, these thinkers provide a multi-dimensional lens for understanding gaze and its implications in HCI and AI systems. Their work guides our effort to mathematically model gaze pressure, identify linguistic behaviors, and offer pathways for epistemic restoration through AI design.

\section{Methodology}

\subsection{Data Collection}

A selection of anonymized Reddit posts was used as the textual data source. Posts were chosen from public forums involving discussions related to marginalized identities and emotional expression. Specifically, \texttt{r/ainbow} (LGBTQ+ identity community) and \texttt{r/conservative} sub-reddits, these sub-reddits represent safe spaces for two communities with different ideologies to derive working results of the model. No endorsement of the platform or posts is implied. These datasets reflect public discourse and was used for non-commercial, research purposes.

\begin{table}[htbp]
\centering
\caption{Coping expressions}
\renewcommand{\arraystretch}{1.2}

\begin{tabularx}{\textwidth}{|p{6.2cm}|X|}
\hline
\textbf{Idea} & \textbf{Coping Expression (Emotion Marker + Sample Phrase)} \\
\hline
Intelligibility of hetero-norms & Hedging: \textit{`I guess..', `Maybe I'm wrong.'} \\
\hline
Proscriptive non-hetero-norms & Overapologizing: \textit{`Sorry if this offends..'} \\
\hline
Compulsory binary gender roles & Self-deprecation: \textit{`I know I'm weird...'} \\
\hline
Non-linear absoluteness & Rigid speech: \textit{`Always', `Never'} \\
\hline
Subjectless verbs & Passive voice: \textit{`It happened...'} \\
\hline
Collective identity & Overcompensation: \textit{`I'll work harder'} \\
\hline
Class reframing & Deflecting: \textit{`It's about class too...'} \\
\hline
Collectivity/systemic words & Generalizing: \textit{`People like us...'} \\
\hline
Male positive framing & Boasting: \textit{`I'm just confident'} \\
\hline
Female negative framing & Negative self-talk \\
\hline
Active/passive roles & Hiding identity: \textit{`I'm just different...'} \\
\hline
Owning/owned dynamic & Isolation narrative: \textit{`I didn't fit in...'} \\
\hline
Speaking/spoken for & Disillusionment: \textit{`No system works...'} \\
\hline
Stigmatization & Hopelessness: \textit{`They never helped...'} \\
\hline
Power imbalance \& isolation & Cynicism: \textit{`They just want a token..'} \\
\hline
Political intersectionality & Resignation: \textit{`That's just how it is..'} \\
\hline
Representational intersectionality & Confusion: \textit{`I don't know who I am..'} \\
\hline
Interlocking systems & Silencing self: \textit{`Why bother speaking?'} \\
\hline
Identity fragmentation/compression & Numbness: \textit{`Whatever..'} \\
\hline
Identity + power & Emotional fatigue: \textit{`I'm tired of explaining...'} \\
\hline
Alien factor & Dismissal: \textit{`It's not worth it..'} \\
\hline
Emotional labor and killjoys & Appeasing language: \textit{`Let's be neutral...'} \\
\hline
Political emotional circulation & Intellectualizing: \textit{`It's not academic...'} \\
\hline
Mediated voice & Disengagement: \textit{`Not part of my world..'} \\
\hline
\raggedright Sanctioned ignorance & Dismissiveness: \textit{`It's just the way it is'} \\
\hline
\raggedright Othering by Enlightenment & Derision: \textit{`Primitive thinking...'} \\
\hline
\raggedright Implied privilege & Denial or Deflection: \textit{`It's not about race'} \\

\hline
\end{tabularx}
\end{table}

\subsection{Annotation Approaches}
A set of coping behavior categories was derived through close review of interdisciplinary literature, including work by Judith Butler, Monique Wittig, Erving Goffman, Sara Ahmed, 
and Gayatri Spivak, among others. Table 1: Coping expressions are mapped to conceptual ideas 
from critical theory; maps key theoretical constructs to linguistic coping manifestations. These manifestations were expressed both emotionally and structurally, providing a foundation for tagging downstream NLP datasets.

\section{Methodology: GPI-Diff – Quantifying Epistemic Divergence through Coping Behavior}

I present a triangulated framework to assess emotional and epistemic divergence between groups using linguistic coping markers. Our approach uses 28 empirically and therapeutically relevant coping labels derived from zero-shot classification (e.g., \textit{hedging}, \textit{over-apologizing}, \textit{disengagement}) to generate structured behavioral profiles.

\subsection{Dimensional Normalization}

Each dataset was z-score normalized across all coping dimensions, allowing for comparison of relative prominence rather than raw magnitude. This normalization supports detection of divergence in emotional \textit{style}.

\subsection{GPI-Diff Composite Metric}

I compute a composite gaze pressure index difference (GPI-Diff) using three components:
\begin{enumerate}
    \item \textbf{Cosine Distance} (\( \text{Cosine} \)) between mean z-normalized coping vectors, representing directional divergence in coping emphasis.
    \item \textbf{Eigenvalue Shift} (\( \text{EigenShift} \)) from Principal Component Analysis (PCA), measuring variance structure divergence across dimensions.
    \item \textbf{Euclidean Distance} (\( \text{Euclidean} \)) of mean raw coping profiles, reflecting overall intensity of emotional effort.
\end{enumerate}

I normalize EigenShift by \( \sqrt{n} \) where \( n = 28 \) (coping dimensions), and compute the harmonic mean of Cosine and EigenShift. The final GPI-Diff is defined as:

\[
\text{GPI-HM}(A, B) = \text{HarmonicMean}\left( \text{Cosine}(A,B), \frac{\text{EigenShift}(A,B)}{\sqrt{n}} \right)
\]

\[
\text{GPI-Diff}(A, B) = \text{GPI-HM}(A, B) + \text{Euclidean}(A, B)
\]

\vspace{1em}

\section{Results}

I applied this metric to 100 Reddit comments each from two ideologically distinct subreddits—\texttt{r/ainbow} (LGBTQ+ identity community) and \texttt{r/conservative} (right-wing political discourse). The following results were obtained:

\begin{itemize}
    \item \textbf{Cosine Distance (directional coping divergence):} 1.0000
    \item \textbf{Eigenvalue Shift (coping structure divergence):} 0.0038
    \item \textbf{Euclidean Distance (coping effort magnitude):} 0.7425
    \item \textbf{Harmonic Mean (Cosine + Eigen):} 0.0076
    \item \textbf{Final GPI-Diff Score:} \textbf{0.7501}
\end{itemize}

\vspace{1em}

\section{Conclusion}

These results reveal that while the two groups exhibit almost identical internal variance structures in how emotional labor is distributed (eigen shift \( \approx 0.0038 \)), they differ entirely in the \textit{style} of coping used (cosine = 1.0), and express a substantial difference in emotional burden (euclidean = 0.7425). The resulting composite GPI-Diff score of 0.7501 demonstrates a robust epistemic and affective divergence between the groups.

This divergence justifies the need for gaze-aware computational systems and psychologically responsive interface design—especially in mixed-group or cross-cultural environments. John Medina stresses on Neuro-plastic abilities practiced through spaced repetitions and exposure, training LLMs on this equation does just that. Such systems can foster epistemic spaciousness by recognizing emotion-linked communication constraints and offering real-time linguistic, emotional, and structural feedback—ultimately enabling more inclusive and accurate social cognition across domains  \cite{medina2014brain}.

\appendix
\section*{Supplementary Material}

A reproducible implementation of the GPI-Diff pipeline, including data ingestion, Hugging Face zero-shot classification, z-score normalization, and final index scoring, is available in the accompanying Jupyter notebook \texttt{GPI\_Diff\_With\_HuggingFace.ipynb}.

\end{document}